\documentclass[twocolumn,showpacs,preprintnumbers,amsmath,amssymb]{revtex4}
\usepackage{graphicx}
\usepackage{dcolumn}
\usepackage{bm}
\usepackage{epsfig}

\begin{document}
\title{Diffusion-controlled formation and collapse of a $d$-dimensional $A$-particle island in the $B$-particle sea}
\author{ Boris ~M.~Shipilevsky}

\affiliation {Institute of Solid State Physics, Chernogolovka,
Moscow district, 142432, Russia\\}

\date{\today}

\begin{abstract}
We consider diffusion-controlled evolution of a $d$-dimensional
$A$-particle island in the $B$-particle sea at propagation of the
sharp reaction front $A+B\to 0$ at equal species diffusivities.
The $A$-particle island is formed by a localized (point)$A$-source
with a strength $\lambda$ that acts for a finite time $T$. We
reveal the conditions under which the island collapse time $t_{c}$
becomes much longer than the injection period $T$ (long-living
island) and demonstrate that regardless of $d$ the evolution of
the long-living island radius $r_{f}(t)$ is described by the
universal law $\zeta_{f}=r_{f}/r_{f}^{M}=\sqrt{e\tau|\ln\tau|}$
where $\tau=t/t_{c}$ and $r_{f}^{M}$ is the maximal island
expansion radius at the front turning point $t_{M}=t_{c}/e$. We
find that in the long-living island regime the ratio $t_{c}/T$
changes with the increase of the injection period $T$ by the law
$\propto (\lambda^{2}T^{2-d})^{1/d}$ i.e. increases with the
increase of $T$ in the one-dimensional (1D) case, does not change
with the increase of $T$ in the 2D case and decreases with the
increase of $T$ in the 3D case. We derive the scaling laws for
particles death in the long-living island and determine the limits
of their applicability. We demonstrate also that these laws
describe asymptotically the evolution of the $d$-dimensional
spherical island with a uniform initial particle distribution
generalizing the results obtained earlier for the
quasi-one-dimensional geometry. As striking results we present a
systematic analysis of the front relative width evolution for
fluctuation, logarithmically modified and mean-field regimes and
demonstrate that in a wide range of parameters the front remains
sharp up to a narrow vicinity of the collapse point.
\end{abstract}
%\pacs{05.70.Ln, 82.20.-w}
%]                                                 %twocolumn format

\maketitle

\narrowtext

\section {Introduction}
For the last decades the reaction-diffusion systems $A + B
\rightarrow 0$, where unlike species $A$ and $B$ diffuse and
irreversibly annihilate in the bulk of a $d$-dimensional medium,
have attracted great interest owing to the remarkable property of
effective {\it dynamical repulsion} of unlike species (
\cite{kbr}-\cite{cot}). In systems with initially spatially
separated reactants this property results in the formation and
self-similar propagation of a {\it localized reaction front}
which, depending on the interpretation of $A$ and $B$ (chemical
reagents, quasiparticles, topological defects, etc.), plays a key
role in a wide range of applications from Liesegang patterns
formation \cite{ant},\cite{rz}, \cite{tho} to electron-hole
luminescence in quantum wells \cite{but1}, \cite{snoke},
\cite{but2}. The simplest model of a planar reaction front,
introduced by Galfi and Racz (GR) \cite{gal}, is the
quasi-one-dimensional model
\begin{eqnarray}
\partial a/\partial t = D_{A}\nabla^{2} a - R, \quad
\partial b/\partial t = D_{B}\nabla^{2} b - R
\end{eqnarray}
for two initially separated reactants which are uniformly
distributed on the left side ($x< 0$) and on the right side ($x>
0$) of the initial boundary. Taking the reaction rate in the
mean-field form $R(x,t)=ka(x,t)b(x,t)$ ($k$ being the reaction
constant), GR discovered that in the long time limit $kt\gg 1$ the
reaction profile $R(x,t)$ acquires a universal scaling form with
the width $w\propto (t/k^{2})^{1/6}$ so that on the diffusion
length scale $L_{D}\propto t^{1/2}$ the relative width of the
front asymptotically contracts unlimitedly $w/L_{D}\sim
(kt)^{-1/3}\to 0$ as $kt\to\infty$. Based on this fact a general
concept of the front dynamics, the quasistatic approximation
(QSA), was developed \cite{ben}- \cite{koza}. The key property of
the QSA is that $w(J)$ depends on $t$ only through the
time-dependent boundary current, $J_{A}=|J_{B}|=J$, the
calculation of which is reduced to solving the external diffusion
problem with the moving {\it absorbing boundary} (Stefan problem)
$R=J\delta(x-x_{f})$. Following this approach, in most subsequent
works the use of the QSA was traditionally restricted by the GR
{\it sea-sea} problem with unlimited number of $A$ and $B$
particles where the stage of monotonous quasistatic front
propagation is always reached asymptotically.

In the recent series of works \cite{self1}-\cite{self5} it has
been shown that based on the QSA the scope of the $A+B\to 0$
problems which allow for analytic description can be appreciably
broadened including the systems with {\it finite} number of
particles and {\it nonmonotonous} front propagation where
asymptotically the QSA is violated. In the Rapid Communication
\cite{self1} the problem of the death of an $A$-particle island in
the uniform $B$-particle sea at equal species diffusivities was
considered.It has been established, that at sufficiently large
initial number of $A$ particles, $N_{0}$, and a sufficiently large
reaction constant $k$ the death of majority of island particles
$N(t)$ proceeds in the universal scaling regime
\begin{eqnarray}
N=N_{0}{\cal G}(t/t_{c}),
\end{eqnarray}
where $t_{c}\propto N_{0}^{2}$ is the lifetime of the island in
the limit $k, N_{0}\to\infty$. This result was obtained under the
assumption that the reaction front propagates quasistatically and
is sustained to be quite sharp, $w/x_{f}\ll 1$ until the collapse
time $t\approx t_{c}$, so that the law of the motion of the front
center $x_{f}(t)$ governs the island width evolution. It has been
shown that while dying, the island first expands to a certain
maximal amplitude $x_{f}^{M}\propto N_{0}$ and then begins to
contract by the law
\begin{eqnarray}
x_{f}=x_{f}^{M}\zeta_{f}(t/t_{c}),
\end{eqnarray}
so that on reaching the maximal expansion amplitude $x_{f}^{M}$
(the turning point of the front),
\begin{eqnarray}
t_{M}/t_{c}=1/e, \quad N_{M}/N_{0}=0.19886...,
\end{eqnarray}
and, therefore, irrespective of the initial particle number,
$\approx 4/5$ of the particles die at the stage of the island
expansion and the remaining $\approx 1/5$ at the stage of its
subsequent contraction. According to \cite{self1} the rapid island
contraction is accompanied by the rapid growth of the front width
$w$ and, therefore, in some vicinity of the collapse point ${\cal
T}=(t_{c}-t)/t_{c}\sim {\cal T}_{Q}$ the reaction front becomes
"blurred" $(w/x_{f}\sim 1)$ and the QSA is no longer applicable.
In the work \cite{self1} it has been shown that for the mean-field
front at small ${\cal T}$
\begin{eqnarray}
w/x_{f}\propto ({\cal T}_{Q}/{\cal T})^{2/3},
\end{eqnarray}
where
$$
{\cal T}_{Q}\propto 1/N_{0}\sqrt{k}
$$
so that ${\cal T}_{Q}\to 0$ at large $k, N_{0}\to\infty$.

It should be emphasized, however, that as well as in the case of
GR problem these results have been obtained for
quasi-one-dimensional geometry (flat front), whereas in numerous
applications the urgent need of their generalization for the
islands possessing circular (ring-shaped front) or spherical
(spherical front) symmetry arises \cite{fi}, \cite{racz2},
\cite{wi}. Moreover, in the paper \cite{self1} the evolution of
the formed island with uniform initial distribution of $A$
particles has been regarded, whereas in many applications the
$A$-particle island appears in the uniform $d$-dimensional $B$
particle sea from a localized (point) source that forms
asymptotically the $d$-dimensional radially symmetric island with
particle distribution dependent on intensity and duration of the
source action (electron-hole luminescence in quantum wells with
localized laser injection of holes into a uniform electron sea is
a prominent illustration of such systems \cite{but1},
\cite{snoke}, \cite{but2}). The regularities of the
$d$-dimensional radially symmetric island growth from a continuous
in time localized (point) source for the physically most important
situation when both $A$ and $B$ particles are mobile were studied
in the work \cite{self2}. In the assumption of sharp reaction
front formation (QSA) it was established that in the
one-dimensional (1D) case the island grows unlimitedly at any
reduced source strength $\lambda $, and the dynamics of its growth
does not depend asymptotically on the diffusivity of $B$
particles. In the 3D case the island grows only at
$\lambda>\lambda _{c}$, achieving asymptotically a stationary
state (static island). In the marginal 2D case the island grows
unlimitedly at any $\lambda $ but at $\lambda <\lambda _{\ast }$
the time of its formation becomes exponentially large.

In this paper, alongside with generalization of the results
\cite{self1} for the $d$-dimensional sphere, as the main goal we
study in the frameworks of the QSA the regularities of evolution
and collapse of the $d$-dimensional $A$-particle island after
switching-off a localized source acting for some finite time $T$.
In the assumption of sharp reaction front formation we reveal the
complete picture of evolution of front trajectory and particle
distribution in the island at change in intensity and duration of
the source action at equal species diffusivities. We focus chiefly
on the situation when the island collapse time $t_{c}$ becomes
much longer than the injection period $T$ (long-living island) and
demonstrate that, in qualitative contrast to a radical change in
the island growth laws \cite{self2} at the change in the system
dimension, the evolution of the long-living island radius
regardless of $d$ is described by the universal law
$$
\zeta_{f}= r_{f}/r_{f}^{M}=\sqrt{e\tau|\ln\tau|},
$$
where $\tau =t/t_{c}$ and $r_{f}^{M}$ is the radius of the island
maximal expansion at the front turning point $t_{M}=t_{c}/e$. We
reveal the conditions of long-living island formation and discover
that in the long-living island regime the ratio $t_{c}/T$ changes
at the increase in the injection period $T$ by the law $\propto
(\lambda ^{2}T^{2-d})^{1/d}$ i.e. it increases at the increase of
$T$ in the 1D case, does not change at the increase of $T$ in the
2D case and decreases at the increase of $T$ in the 3D case. We
show that in the long-living island regime by the time moment of
source switching-off the majority of injected particles survives
and derive scaling laws of particle death in the island. We also
demonstrate that asymptotically these laws describe the evolution
of the $d$- dimensional spherical island with uniform initial
particle distribution. Finally, we analyze self-consistently the
regularities of the reaction front relative width evolution for
fluctuation, logarithmically modified and mean-field regimes.

\section {Evolution of the $d$-dimensional spherical island with uniform initial particle distribution}

We start with generalization of the paper \cite{self1} results
obtained for quasi-one-dimensional geometry. Let the uniform
$d$-dimensional spherically symmetrical $A$-particle island with a
radius $L$ ($r\in [0,L)$) be "submerged" into the uniform
$d$-dimensional sea of particles $B$ $(r\in (L,\infty ))$ with
initial concentrations $a_{0}$ and $b_{0}$, respectively (it is
clear that in the 1D case the interval $r\in[0, \infty)$
represents a half of radially symmetric distribution $r=|x|$).
Particles $A$ and $B$ diffuse with nonzero diffusion constants
$D_{A,B}$ and upon contact annihilate with some nonzero
probability, $A+B\to 0$. In the continuum version this process can
be described by the reaction-diffusion equations (1) where
$a(r,t)$ and $b(r,t)$ are the mean local concentrations of $A$ and
$B$ which, by symmetry, we  assume to be dependent only on the
radius, and $R(r,t)$ is the macroscopic reaction rate. We shall
assume, as usual, that species diffusivities are equal
$D_{A}=D_{B}=D$. This important condition, due to local
conservation of difference concentration $a-b$, leads to a radical
simplification that permits to obtain an analytical solution for
arbitrary front trajectory (we shall note that at different
species diffusivities $D_{A}\neq D_{B}$ an analytical solution of
the Stefan problem is possible only for a stationary or a
monotonically moving front \cite{self5}). Then, by measuring the
length, time and concentration in units of $L, L^{2}/D$, and
$b_{0}$, respectively, and defining the ratio $a_{0}/b_{0}=c$, we
come from Eq.(1) to the simple diffusion equation for the
difference concentration $s(r,t)=a(r,t)-b(r,t)$
\begin{eqnarray}
\partial s/\partial t = \nabla^{2} s,
\end{eqnarray}
at the initial conditions
\begin{eqnarray}
s_{0}(r\in [0,1))=c, \quad s_{0}(r\in (1,\infty])=-1,
\end{eqnarray}
with the boundary conditions
\begin{eqnarray} \nabla
s\mid_{r=0}=0, \quad s(\infty, t)= -1.
\end{eqnarray}
As well as in the paper \cite{self1} we shall assume that the
ratio of concentrations island/sea is large enough, $c\gg 1$(
concentrated island). Below it will be shown that in the limit of
large $c\gg 1$ the "lifetime" of the island $t_{c}\propto
c^{2/d}\gg 1$, so the majority of the particles die at times $t\gg
1$, when the diffusive length exceeds appreciably the initial
island radius. As well as in the paper \cite{self1} the evolution
of the island in such a large-$t$ regime is of principal interest
to us here.

Asymptotics of the exact solution of the problem (6)-(8) for
$d=1,2,3$ at large $t$ and $r/t\ll 1$ has the form
\begin{eqnarray}
s(r,t)=\frac{\gamma N_{0}}{(4\pi
t)^{d/2}}e^{-r^{2}/4t}(1-\chi_{d})-1,
\end{eqnarray}
where $N_{0}=\mu_{d}c (\mu_{1}=2, \mu_{2}=\pi, \mu_{3} =4\pi/3)$
is the initial number of particles in the island in units of
$b_{0}L^{d}$, $\gamma=(c+1)/c$ and
$$
\chi_{d}=\alpha_{d}(1-r^{2}/2dt)/t+\cdots
$$
with $\alpha_{1}=1/12$, $\alpha_{2}=1/8$ and $\alpha_{3}=3/20$.
According to the QSA for large $k\to\infty$ at times $t\propto
k^{-1}\to 0$ there forms a sharp reaction front $w/r_{f}\to 0$ so
that in neglect of the reaction front width the solution $s(r,t)$
defines the law of its propagation $s(r_{f},t)=0$ and the
evolution of particle distributions $a=s (r<r_{f})$ and $b=|s|
(r>r_{f})$. Substituting to Eq.(9) the condition $s(r_{f},t)=0$
and assuming that $\chi _{d}^{f}\ll 1$ we find the law of the
front motion in the form
\begin{eqnarray}
r_{f}=2(1+\alpha_{d}/dt)\sqrt{t\ln \left[\frac{\gamma
N_{0}(1-\alpha_{d}/t)}{(4\pi t)^{d/2}}\right]}.
\end{eqnarray}
From Eq.(10) it follows that at any $d$ in the limit of large
$t,N_{0}\gg 1$ the island first expands reaching some maximal
radius $r_{f}^{M}$ , and then it contracts disappearing in the
collapse point $t_{c}\propto N_{0}^{2/d}$. Taking $r_{f}(t_{c})=0$
we find from Eq. (10)
\begin{eqnarray}
t_{c}=\frac{(\gamma N_{0})^{2/d}[1-{\cal O}(N_{0}^{-2/d})]}{4\pi},
\end{eqnarray}
whence neglecting the terms ${\cal O}[{\rm max}(1/c, 1/c^{2/d}])$
we obtain
\begin{eqnarray}
t_{c}=(N_{0})^{2/d}/4\pi.
\end{eqnarray}
Neglecting further the terms $\alpha_{d}/t$ we finally find from
Eq. (10)
\begin{eqnarray}
r_{f}=\sqrt{2dt\ln(t_{c}/t)},
\end{eqnarray}
whence it follows immediately that in the front turning point
$\dot{r}_{f}(t_{M})=0$
\begin{eqnarray}
t_{c}/t_{M}=e
\end{eqnarray}
and
\begin{eqnarray}
r_{f}^{M}=\sqrt{2dt_{M}}=(N_{0})^{1/d}\sqrt{d/2\pi e}.
\end{eqnarray}
Introducing the scaling variables $\zeta =r/r_{f}^{M}$ and $\tau
=t/t_{c}$ we come to the result announced above that in the limit
of large $t,N_{0}\gg 1$ {\it regardless} of the system dimension
the front trajectory is described by the universal law
\begin{eqnarray}
\zeta_{f}(\tau)=r_{f}/r_{f}^{M}=\sqrt{e\tau|\ln\tau|}.
\end{eqnarray}
It should be noted that taking into account Eq. (13) in the limit
of large $t_{c}$ the condition $\chi _{d}^{f}\ll 1$ reduces to the
more rigid requirement $t\gg \alpha _{d}\ln (t_{c}/t)$. Assuming
that this condition is fulfilled and neglecting the reaction front
width, at the same approximation as above for particle
distribution in the island we find from Eq. (9)
\begin{eqnarray}
a(\zeta, \tau)= s(\zeta, \tau)=(e^{-\zeta^{2}/e\tau}/\tau)^{d/2} -
1.
\end{eqnarray}
Calculating further the number of particles in the island
$N=g_{d}(r_{f}^{M})^{d}\int_{0}^{\zeta_{f}}a(\zeta, \tau)
\zeta^{d-1}d\zeta$ (here $g_{1}=2, g_{2}=2\pi$ and $g_{3}=4\pi$)
we obtain immediately the scaling laws of particles death in the
island
\begin{eqnarray}
N=N_{0}{\cal G}_{d}(\tau),
\end{eqnarray}
where
$$
{\cal G}_{1}(\tau)={\rm erf}(\sqrt{|\ln \tau|/2})-\sqrt{2\tau |\ln
\tau|/\pi},
$$
$$
{\cal G}_{2}(\tau)=1-\tau(1+|\ln \tau|),
$$
$$
{\cal G}_{3}(\tau)={\rm erf}(\sqrt{3|\ln
\tau|/2})-\sqrt{6\tau^{3}|\ln \tau|/\pi}(1+|\ln \tau|).
$$
From Eqs. (17) and (18) we conclude that in the front turning
point $\tau_{M}=1/e$ {\it regardless} of the initial number of $A$
particles their concentration in the center of the island
\begin{eqnarray}
a(0,\tau_{M})=e^{d/2}-1
\end{eqnarray}
and the fraction of the particles that survived in the process of
the island expansion
\begin{eqnarray}
N_{M}/N_{0}= \left\{\begin{array} {lcl} 0.19886..., \quad
d=1,\\
0.26412..., \quad d=2,\\
0.29986..., \quad d=3.\\
\end{array}\right.
\end{eqnarray}
Assuming that the front remains sharp enough up to a narrow
vicinity of the collapse point ${\cal T}=(t_{c}-t)/t_{c}\ll 1$ we
find from Eqs.(16) and (18) that at small ${\cal T}\ll 1$ at the
final collapse stage particles death proceeds by the law
\begin{eqnarray}
N/N_{0}=c_{d}{\cal T}^{(d+2)/2}=c_{d}(\zeta_{f}/\sqrt{e})^{d+2},
\end{eqnarray}
where $c_{1}=\sqrt{2/\pi}/3, c_{2}=1/2$ and $c_{3}=
3\sqrt{6/\pi}/5$. In Fig.1 are shown the dependencies $\zeta
_{f}(\tau )$ and ${\cal G}_{d}(\zeta _{f})$ that demonstrate the
key features of evolution of 1D, 2D and 3D islands in the limit of
large $N_{0}\rightarrow \infty $.

One of the central points of our analysis is revealing of
applicability limits for the assumption that the formed reaction
front remains sharp enough, $\eta =w/r_{f}\ll 1$, up to a narrow
vicinity of the collapse point. The detailed discussion of this
problem will be presented in Section IV. Completing this section
we shall reveal the regularities for the evolution of the boundary
current density $J=-\partial a/\partial r|_{r=r_{f}}$ that
according to the QSA determines the evolution of the reaction
front width $w(J)$. From Eqs. (16) and (17) we find easily
\begin{eqnarray}
{\cal J}(\tau)=J/J_{M}=\sqrt{\frac{|\ln\tau|}{e\tau}},
\end{eqnarray}
where $J_{M}=d/r_{f}^{M}=\sqrt{2\pi ed}/(N_{0})^{1/d}$. Thus, we
conclude that as well as the front trajectory (16) the boundary
current evolution is described by the universal law (22)
regardless of the system dimension that predetermines the
universality of the mean-field front relative width evolution.

\section {Evolution of the $d$-dimensional island formed by a localized source}

Let us proceed now to the analysis of formation regularities of
the $d$-dimensional $A$-particle spherical island from a localized
(point) $A$-particle source and the consequent island evolution
after switching-off this source. Let particles $A$ be injected at
$t\geq 0$ with a rate $\Lambda$ at the point ${\bf r}=0$ of the

\begin{figure}
\includegraphics[width=1\columnwidth]{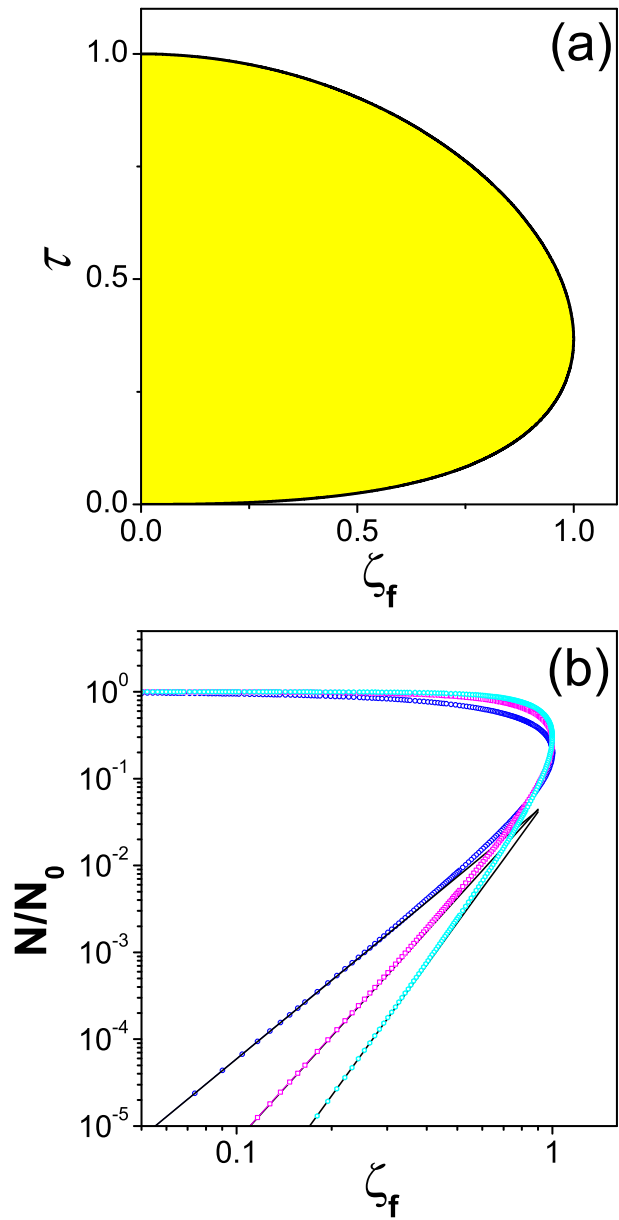}
\caption {(Color online) (a) Universal trajectory of the reduced
front radius $\zeta_{f}(\tau)$ Eq.(16) (line). The island area is
colored. (b) Scaling functions $G_{d}(\zeta_{f})=N/N_{0}$
calculated according to Eqs. (16) and (18) for $d=1$(circles),
$d=2$(squares) and $d=3$(hexagons). Straight lines show the
small-$\zeta_{f}$ asymptotes according to Eqs. (21).}
\label{fig 1}
\end{figure}

uniform $d$-dimensional sea of particles $B$, distributed with a
density $\rho$. The source is acting for some finite time $T$, and
then it is switched-off. As above, particles $A$ and $B$ diffuse
with nonzero diffusion constants $D_{A,B}=D$ and upon contact
annihilate with some nonzero probability, $A+B\to 0$. In the
continuum version this process can be described by the
reaction-diffusion equations (1) with an additional source term
$\Lambda \delta({\bf r})\Theta(T-t)\Theta(t)$ for particles $A$
($\Theta$ denotes here the Heaviside step function) where $a(r,t)$
and $b(r,t)$, by symmetry, we assume to be dependent only on the
radius with the initial conditions $a(r,0)=0, b(r,0)=\rho$, and
the boundary condition $b(\infty,t)=\rho$. The initial density of
the sea, $\rho$, defines a natural scale of concentrations and a
characteristic length scale of the problem - the average
interparticle distance $\ell=\rho^{-1/d}$. So, by measuring the
length, time and concentration in units of $\ell, \ell^{2}/D$, and
$\rho$, respectively, we introduce the dimensionless source
strength $\lambda=\Lambda \ell^{2}/D$ and the dimensionless
reaction constant $\kappa= k\rho\ell^{2}/D$. Defining then the
difference concentration $s(r,t)=a(r,t)-b(r,t)$ we come to the
simple diffusion equation with source
\begin{eqnarray}
\partial s/\partial t = \nabla^{2} s + \lambda \delta({\bf r})\Theta(T-t)\Theta(t),
\end{eqnarray}
at the initial and boundary conditions
\begin{eqnarray}
s(r,0)=s(\infty, t)=-1.
\end{eqnarray}
According to Eq. (23) in the course of injection in the vicinity
of the source there arises a region of $A$-particle excess,
$s(r,t)>0$, which expands with time. Following the paper
\cite{self2} we shall assume that, by analogy with the Galfi-Racz
problem, a narrow reaction front has to form at this region
boundary, for which the law of motion, $r_{f}(t)$, according to
the QSA, can be derived from the condition $s(r_{f},t)=0$.

The exact solution of the problem (23), (24) at the injection
stage $0<t\leq T$ has the form
\begin{eqnarray}
s+1= \frac{\lambda}{(4\pi)^{d/2}}\int_{0}^{t} d\theta
e^{-r^{2}/4(t-\theta)}/(t-\theta)^{d/2},
\end{eqnarray}
whereas after source switching-off at $t>T$ we find
\begin{eqnarray}
s+1= \frac{\lambda}{(4\pi)^{d/2}}\int_{0}^{T} d\theta
e^{-r^{2}/4(t-\theta)}/(t-\theta)^{d/2}.
\end{eqnarray}
We shall start with the discussion of the key features of the
$d$-dimensional island formation and growth at the injection stage
reproducing here partially the paper \cite{self2} results for the
completeness.

\subsection{Formation and growth of the $d$-dimensional $A$-particle island at
the injection stage $0<t\leq T$}

Integrating Eq. (25) for $d=1,2,3$ we find immediately
\begin{eqnarray} s+1 = \sqrt{\lambda^{2}t} {\rm
ierfc}\left(\frac{r}{2\sqrt{t}}\right), \quad d=1,
\end{eqnarray}
\begin{eqnarray}
s+1=-(\lambda/4\pi){\rm Ei}\left(-\frac{r^{2}}{4t}\right), \quad
d=2,
\end{eqnarray}
\begin{eqnarray}
s+1= \left(\frac{\lambda}{4\pi r}\right){\rm
erfc}\left(\frac{r}{2\sqrt{t}}\right), \quad d=3,
\end{eqnarray}
whence, assuming the front to be formed ($w/r_{f}\ll 1$) and
neglecting its width, from the condition $s(r_{f},t)=0$ we obtain
the laws for the island radius (the front center radius) growth
$r_{f}(t)$
\begin{eqnarray}
{\rm ierfc}(r_{f}/2\sqrt{t})= 1/\sqrt{\lambda^{2}t}, \quad d=1,
\end{eqnarray}
\begin{eqnarray}
{\rm Ei}(- r_{f}^{2}/4t)= -4\pi/\lambda, \quad d=2,
\end{eqnarray}
\begin{eqnarray}
{\rm erfc}(r_{f}/2\sqrt{t})= 4\pi r_{f}/\lambda, \quad d=3,
\end{eqnarray}
where ${\rm ierfc}(u)= \int_{u}^{\infty}{\rm
erfc}(v)dv=e^{-u^{2}}/\sqrt{\pi}-u{\rm erfc}(u)$ and ${\rm
Ei}(-u^{2})= - \int_{u^{2}}^{\infty} dve^{-v}/v$ is the
exponential integral. Calculating further the number of $A$
particles surviving in the injection process
\begin{eqnarray}
N=g_{d}\int_{0}^{r_{f}}s(r,t)r^{d-1}dr
\end{eqnarray}
we conclude from Eqs.(27)-(33) that at $d\neq 2$ the growth of the
island radius $r_{f}(t)$ and the number of surviving $A$ particles
$N(t)$ are described by the scaling laws
\begin{eqnarray}
r_{f}= \lambda^{d-2}{\cal R}_{d}(\lambda^{2/(2-d)}t),
\end{eqnarray}
\begin{eqnarray}
N= \lambda^{d/(d-2)}{\cal N}_{d}(\lambda^{2/(2-d)}t).
\end{eqnarray}
As a consequence, the fraction of surviving $A$ particles is
described by the scaling law
\begin{eqnarray}
q=N/\lambda t= {\cal Q}_{d}(\lambda^{2/(2-d)}t).
\end{eqnarray}
Following the paper \cite{self2} below we focus on the key
features of the $d$-dimensional island growth for each dimension
separately.

\subsubsection{One-dimensional island}

In the 1D case from Eq. (27) it follows that an excess of $A$
particles in the source vicinity forms in a time
$t_{\star}=\pi/\lambda^{2}$ ($s(0,t_{\star})=0$). It is, however,
clear that a hydrodynamic approximation comes into play at times
$t\gg {\rm max}(1, 1/\lambda)$; therefore at early island
formation stages one can distinguish two qualitatively different
island growth regimes: a) $\lambda\ll 1$, when the island
formation proceeds under conditions of death of the majority of
injected particles, and b) $\lambda\gg 1$, when a multiparticle
"cloud" forms long before the beginning of noticeable annihilation
and, therefore, the stage of the developed reaction, $t\gg 1 (\gg
t_{\star})$, is preceded here by a stage of purely diffusive
expansion of the cloud, $1/\lambda\ll t\ll 1$ \cite{self2}. In the
long-time limit $t\gg {\rm max}(1,t_{\star})$ at any $\lambda$ we
obtain from Eqs. (27), (30) and (33) the exact asymptotics
\begin{eqnarray}
r_{f}=\sqrt{2t\ln \Gamma}(1-\ln\ln \Gamma/\ln \Gamma +\cdots),
\end{eqnarray}
\begin{eqnarray}
N=\lambda t[1-{\cal O}(\sqrt{\ln\Gamma/\Gamma})],
\end{eqnarray}
where $\Gamma = t/t_{\star}$. So, we conclude that forming in
qualitatively different regimes from $q\ll 1 (\lambda\ll 1)$ to
$1-q\ll 1 (\lambda\gg 1)$ the 1D island at any $\lambda$ crosses
over to the universal growth regime (37), (38) with an unlimited
decay of the dying particle fraction $1-q\propto
\sqrt{\ln\Gamma/\Gamma}\to 0$ as $\Gamma\to\infty$. Fig.2 shows
the dependencies $\lambda r_{f}$ vs $\lambda ^{2}t$ calculated
according to Eqs. (30) and (37). One can see that asymptotics (37)
gives an exact description of the front trajectory starting with
$\lambda ^{2}t\sim 10^{3}$.

\begin{figure}
\includegraphics[width=1\columnwidth]{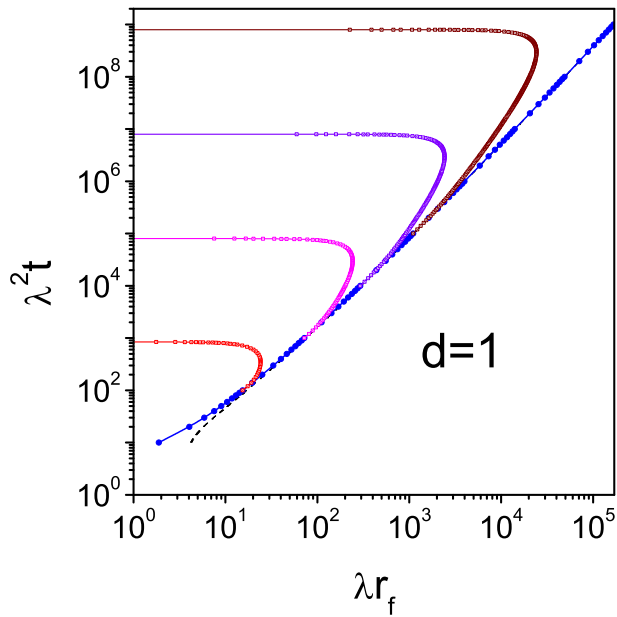}
\caption {(Color online) Filled circles show the trajectory of the front
radius for the continuous in time point source in the
one-dimensional medium calculated according to Eq. (30) in the
scaling coordinates $\lambda r_{f}$ vs. $\lambda^{2}t$. Dashed
line shows long-time asymptotics Eq.(37). Open squares show the
trajectories of the front radius after the source switching-off
calculated from Eq. (44) for the parameter values $\lambda^{2}T =
10^{2}, 10^{3}, 10^{4}$ and $10^{5}$(from bottom to top).}
\label{fig 2}
\end{figure}

\subsubsection{Two-dimensional island}

In the 2D case from Eqs. (28), (31) and (33) we find
\begin{eqnarray}
r_{f}=2\sqrt{\alpha t},
\end{eqnarray}
\begin{eqnarray}
N=\lambda t(1-e^{-\alpha}),
\end{eqnarray}
where $\alpha$ is the root of the equation ${\rm
Ei}(-\alpha)=-\lambda_{*}/\lambda$, $\lambda_{*}=4\pi$ and has the
asymptotics $\alpha=e^{-\lambda_{*}/\lambda}/\gamma$
$(\gamma=1.781...)$ at $\lambda\ll \lambda_{*}$ and
$\alpha=\ln(\lambda/\lambda_{*}\alpha)$ at $\lambda\gg
\lambda_{*}$. We conclude that in 2D the island growth rate
$\alpha$ and the fraction of surviving $A$ particles $q$ {\it do
not vary in time}: at large $\lambda\gg \lambda_{*}$ the majority
of injected particles survive
$$
1-q \sim \frac{\ln\lambda}{\lambda}\ll 1, \quad \lambda\gg
\lambda_{*},
$$
whereas at small $\lambda\ll \lambda_{*}$ the majority of injected
particles die
$$
q\sim e^{-\lambda_{*}/\lambda}\ll 1, \quad \lambda\ll \lambda_{*}.
$$
One of the key consequences of Eqs. (39) and (40) consists in the
exponentially strong decrease of the growth rate $\alpha$ and the
fraction of surviving particles $q$ in the region $\lambda\ll
\lambda_{*}$. It means that though in the 2D case the island
unlimitedly growth at any $\lambda$, at small $\lambda\ll
\lambda_{*}$ the 2D island growth is actually {\it suppressed}.

\subsubsection{Three-dimensional island}

In the 3D case from Eqs. (29) and (32) it follows that at any
$\lambda$ in the long-time limit $t\gg
t_{s}=(\lambda/\lambda_{*})^{2}$ the front radius by the law
$$
r_{f}=r_{s}[1-{\cal O}(\sqrt{t_{s}/t})]
$$
reach a stationary value ({\it stationary island})
\begin{eqnarray}
r_{f}(t/t_{s}\to\infty)=r_{s}=\lambda/\lambda_{*}.
\end{eqnarray}
According to Eqs. (29), (32) and (33) in this limit the number of
surviving particles is
$$
N=(2\pi/3)r_{s}^{3}[1-{\cal O}(\sqrt{t_{s}/t})],
$$
whence is follows that in radical contrast to the 1D case in the
3D case at any injection rate all the injected particles die
asymptotically $q=(1/6)(t_{s}/t)\to 0$ as $t/t_{s}\to\infty$. In
the paper \cite{self2} it is shown that defining a minimal
stationary island through the condition $w_{s}/r_{s}\sim 1$, we
conclude that in the 3D case the island forms only when the
injection rate exceeds a ${\it critical}$ value
$$
\lambda_{c}\sim \lambda_{*}/\sqrt{\kappa}\gg 1.
$$
One of the key consequences of Eq. (32) is that at high injection
rates $\lambda\gg \lambda_{c}$ the stationary state ($t\gg t_{s}$)
is preceded by an intermediate stage $1\ll t\ll t_{s}$ wherein the
island grows by the law
\begin{eqnarray}
r_{f}=\sqrt{2t\ln(t_{s}/t)}[1-\ln(\sqrt{\pi}\omega)/\omega+\cdots],
\end{eqnarray}
where $\omega=\ln(t_{s}/t)$. According to Eqs.(29), (32) and (33),
at this stage the number of surviving particles is
\begin{eqnarray}
N=\lambda t[1-{\cal O}(\omega^{3/2}\sqrt{t/t_{s}})]
\end{eqnarray}
and therefore the majority of injected particles are {\it still
surviving}
$$
1-q\sim \omega^{3/2}\sqrt{t/t_{s}}\ll 1.
$$
In Ref.\cite{self2} are presented the details of formation of the
spherical island (42), (43) from a diffusive cloud which expands
in the absence of reaction and it is established that the formed
front condition $w/r_{f}\ll 1$ is realized at $t\gg
\ln(t_{s}/t)/\kappa$. Fig.3 shows the dependencies $r_{f}/\lambda
$ vs $t/\lambda ^{2}$ calculated according to Eqs. (32) and (42).
One can see that asymptotics (42) gives an exact description of
the front trajectory up to $t/\lambda ^{2}\sim 10^{-5}(t/t_{s}\sim
10^{-3})$.

\begin{figure}
\includegraphics[width=1\columnwidth]{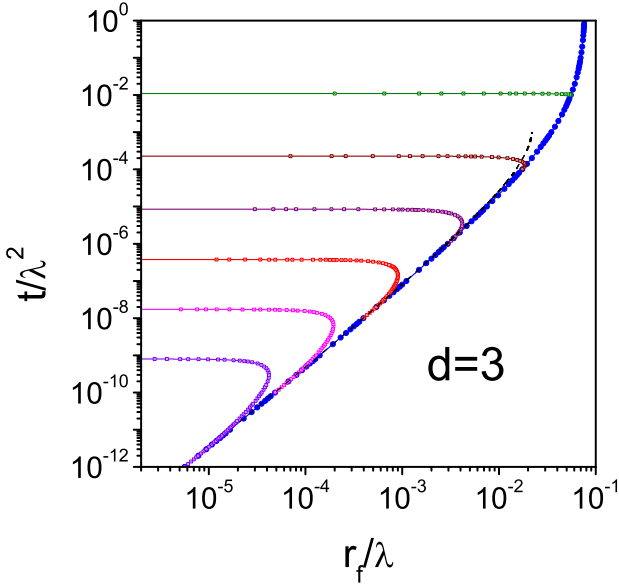}
\caption {(Color online) Filled circles show the  trajectory of the front
radius for the continuous in time point source in the
three-dimensional medium calculated according to Eq. (32) in the
scaling coordinates $r_{f}/\lambda$ vs. $t/\lambda^{2}$. Dashed
line shows the intermediate asymptotics Eq.(42). Open squares show
the trajectories of the front radius after the source switching-off
calculated from Eq. (46) for the parameter values $\lambda^{2}/T =
10^{2}, 10^{4},10^{6}, 10^{8}, 10^{10}$ and $10^{12}$(from top to
bottom).}
\label{fig 3}
\end{figure}

\subsection{Evolution and collapse of the $d$-dimensional island
after source switching-off $t>T$}

According to Eq.(26) we find that at $d=1,2,3$ the evolution of
particles distribution $s(r,t)$ after source switching-off $\Delta
t=t-T>0$ is described by the expressions
\begin{eqnarray}
\nonumber s+1 = \lambda \left[\sqrt{t} {\rm ierfc}
\left(\frac{r}{2\sqrt{t}}\right)-\sqrt{\Delta t} {\rm ierfc}
\left(\frac{r}{2\sqrt{\Delta t}}\right)\right], \quad d=1,
\end{eqnarray}
$$
s+1=\frac{\lambda}{4\pi}[{\rm Ei}(-r^{2}/4\Delta t) - {\rm
Ei}(-r^{2}/4t)], \quad d=2,
$$
$$
s+1= \frac{\lambda}{4\pi r}\left[{\rm
erfc}\left(\frac{r}{2\sqrt{t}}\right) - {\rm
erfc}\left(\frac{r}{2\sqrt{\Delta t}}\right)\right], \quad d=3,
$$
whence neglecting the front width $w/r_{f}\ll 1$ we find from the
condition $s(r_{f},t)=0$ the island radius trajectory $r_{f}(t)$
\begin{eqnarray}
\sqrt{t} {\rm ierfc}
\left(\frac{r_{f}}{2\sqrt{t}}\right)-\sqrt{\Delta t} {\rm ierfc}
\left(\frac{{r}_{f}}{2\sqrt{\Delta t}}\right)=1/\lambda, \quad
d=1,
\end{eqnarray}
\begin{eqnarray}
{\rm Ei}(-r_{f}^{2}/4\Delta t) - {\rm
Ei}(-r_{f}^{2}/4t)=\frac{4\pi}{\lambda}, \quad d=2,
\end{eqnarray}
\begin{eqnarray}
{\rm erfc}\left(\frac{r_{f}}{2\sqrt{t}}\right) - {\rm
erfc}\left(\frac{r_{f}}{2\sqrt{\Delta t}}\right)=\frac{4\pi
r_{f}}{\lambda}, \quad d=3.
\end{eqnarray}
Assuming further that the front remains sharp enough up to a
narrow vicinity of the collapse point $t_{c}$, from the condition
$r_{f}(t_{c})=0 $ we obtain
\begin{eqnarray}
t_{c}=(\lambda T)^{2}(1+\pi/\lambda^{2} T)^{2}/4\pi, \quad d=1,
\end{eqnarray}
\begin{eqnarray}
t_{c}=T/(1-e^{-\lambda_{*}/\lambda}), \quad d=2,
\end{eqnarray}
\begin{eqnarray} 4\pi^{3/2}/\lambda =
1/\sqrt{t_{c}-T}-1/\sqrt{t_{c}}, \quad d=3,
\end{eqnarray}
whence for the ratio of the island collapse time $t_{c}$ to the
injection period $T$ it follows immediately
\begin{eqnarray}
t_{c}/T= {\cal F}_{d}(\lambda^{2}T^{2-d}).
\end{eqnarray}
One can easily be convinced that the asymptotics of the scaling
function ${\cal F}_{d}(z)$ in the limit of large $z\gg 1$ has the
form
\begin{eqnarray}
{\cal F}_{d}(z)= \frac{z^{1/d}(1+2\pi/z^{1/d}+\cdots)}{4\pi}.
\end{eqnarray}
Thus, we conclude that a) at $(\lambda ^{2}T^{2-d})^{1/d}/4\pi \gg
1$ the island lifetime $t_{c}$ becomes much longer than the
injection period $T$ (the long-living island)
\begin{eqnarray}
\nonumber
t_{c}/T = (\lambda^{2}T^{2-d})^{1/d}/4\pi \gg 1
\end{eqnarray}
and b) in the long-living island regime the ratio $t_{c}/T$
increases at the increase of $T$ in the 1D case, does not change
at the increase of $T$ in the 2D case and decreases at the
increase of $T$ in the 3D case. Let us consider the consequences
of Eqs. (44)-(49) for each dimension separately.

\subsubsection{One-dimensional island}

According to Eqs. (37), (38) in the 1D case one of the key
conditions for island formation is the requirement
$\lambda^{2}T\gg 1$ that is why the formed 1D island at {\it any}
$\lambda $ is long-living. From Eq.(44) it follows that after
source switching-off the long-living island continues to expand
reaching the maximum
$$
r_{f}^{M}\propto \lambda T,
$$
whence taking into account Eq. (37) we find
$$
r_{f}^{M}/r_{f}^{T}\propto \sqrt{\frac{\lambda^{2}T}{\ln
{(\lambda^{2}T})}}\to \infty
$$
as $\lambda^{2}T\to\infty$. The trajectories of the 1D island
radius calculated according to Eq. (44) in the scaling coordinates
$\lambda r_{f} $ vs $\lambda ^{2}t$ are shown in Fig.2
demonstrating the evolution of these trajectories with the growing
parameter $\lambda ^{2}T$.

\subsubsection{Two-dimensional island}

According to Eq. (45) in the 2D case in the limit of anomalously
slow growth $\lambda\ll \lambda_{*}$, when the majority of
injected particles die, we find
$$
r_{f}^{M}/r_{f}^{T}\approx 1,
$$
whence taking into account Eq.(48) it follows that regardless of
the injection time after source switching-off the formed 2D island
begins to contract immediately disappearing for exponentially
small (in comparison with the injection period) time interval
$$
(t_{c}-T)/T \sim e^{-\lambda_{*}/\lambda}\ll 1.
$$
In the opposite limit of the long-living island $\lambda\gg
\lambda_{*}$ after source switching-off the island continues to
expand reaching the maximum
$$
r_{f}^{M}\propto \sqrt{\lambda T},
$$
whence taking into account Eq. (39) we find that regardless of the
injection duration
$$
r_{f}^{M}/r_{f}^{T}\propto \sqrt{\frac{\lambda}{\ln\lambda}}\to
\infty
$$
as $\lambda\to\infty$.

\subsubsection{Three-dimensional island}

According to Eq. (46) in the 3D case in the limit of the
stationary island $T\gg t_{s}= (\lambda/\lambda_{*})^{2}$, when
the majority of the injected particles die, we find
$$
r_{f}^{M}/r_{s}\approx 1,
$$
whence, as expected, it follows that in this limit regardless of
the injection duration after source switching-off the formed 3D
island begins to contract immediately disappearing for the time
$t_{c}-T \propto r_{s}^{2}\ll T$. Indeed, from Eq.(49) in the
stationary limit $T\gg t_{s}$ we find
$$
t_{c}-T=\frac{\lambda^{2}(1-\sqrt{\lambda^{2}/T}/2\pi^{3/2}+\cdots)}{16\pi^{3}}.
$$
In the opposite limit of the long-living island $1\ll T\ll t_{s}$
after source switching-off the island continues to expand reaching
the maximum
$$
r_{f}^{M}\propto (\lambda T)^{1/3},
$$
whence taking into account Eq. (42) we find
$$
r_{f}^{M}/r_{f}^{T}\propto \left[\frac
{\lambda^{2}}{T\ln^{3}(t_{s}/T)}\right]^{1/6}\to\infty
$$
as $\lambda^{2}/T\to\infty$. The trajectories of the 3D island
radius calculated according to Eq. (46) in the scaling coordinates
$r_{f}/\lambda$ vs $t/\lambda^{2}$ are shown in Fig.3
demonstrating the evolution of these trajectories with the growing
parameter $\lambda ^{2}/T$.

As it was stated in the introduction and as it is clear from the
presented analysis the evolution regularities in the long-living
island regime are of the main interest that is why below we shall
focus namely on this regime. According to Eqs. (44), (45) and (46)
the maximal volume of $d$-dimensional island expansion in the
front turning point is proportional to the number of particles
injected by the time of source switching-off
$$
\Omega_{M}=\mu_{d}(r_{f}^{M})^{d}\propto N_{T}= \lambda T.
$$
But according to Eqs. (38), (40) and (43) in the long-living
island regime the majority of injected particles survives up to
the time of source switching-off, i.e. the value of $N_{T}$
determines the number of $A$-particles at the moment of source
switching-off. Comparing these results and taking into account
Eqs. (37), (39) and (42) we come to the important conclusion that
in the limit of the long-living island regardless of the medium
dimension and the injection duration the ratio of the
$d$-dimensional island maximal volume in the front turning point,
$\Omega _{M}$, to the starting island volume at the moment of
source switching-off, $\Omega _{T}$, is proportional to the island
particles mean concentration $<a>_{T}=N_{T}/\Omega _{T}$ at the
moment of source switching-off and inversely proportional to the
fraction of the particles died by this moment $1-q_{T}$
\begin{eqnarray}
 <a>_{T}\propto \frac{1}{1-q_{T}}\propto
\frac{\Omega_{M}}{\Omega_{T}}=\left(\frac{r_{f}^{M}}{r_{f}^{T}}\right)^{d}.
\end{eqnarray}
From Eq. (52) it follows that the long-living island regime is
realized in the limit when at the moment of source switching-off
the island particles mean concentration $<a>_{T}$ becomes large as
compared to the initial sea density. The value of $<a>_{T}$
increases with the increase of $T$ in the 1D case, does not change
with the increase of $T$ in the 2D case and decreases with the
increase of $T$ in the 3D case that leads to the corresponding
behavior of the island expansion relative amplitude $\Omega
_{M}/\Omega _{T}$ and, as a consequence, to the corresponding
behavior of  the relative island lifetime $t_{c}/T$.

Comparing the long-living island evolution after source
switching-off to the evolution of the initially uniform
concentrated island we conclude that in both of the problems the
maximal island expansion amplitude $\Omega _{M}$ is determined
unambiguously by the "starting" number of particles in the island
$N_{st}$ and, as a consequence, in both of the problems the island
collapse time is $t_{c}=(N_{st})^{d/2}/4\pi $ where $N_{st}=N_{T}$
or $N_{0}$, respectively. This fact gives grounds to assert that
in the long-time limit ($t\gg T$ and $t\gg 1$, respectively) {\it
regardless} of the starting particle distribution the island
evolution takes a universal form characterizing island death in
the instantaneous source regime. Below we shall demonstrate the
validity of this statement and reveal the conditions for
universalization of the long-time island evolution.

\subsection{Self-similar evolution of the long-living island}

We shall assume that the island lifetime $t_{c}$ exceeds the
injection period $T$ considerably. Then in the long-time limit
$T\ll t<t_{c}$ taking into account the additional requirement
$r^{2}T/t^{2}\ll 1$ from Eq. (26) we easily find the long-time
asymptotics of particle distribution
\begin{eqnarray}
s+1=\frac{\lambda T}{(4\pi
t)^{d/2}}e^{-r^{2}/4t}[1+(d-r^{2}/2t)T/4t +\cdots],
\end{eqnarray}
whence, neglecting the front width, from the condition
$s(r_{f})=0$ we obtain the long-time asymptotics of front
trajectory
\begin{eqnarray}
r_{f}=2(1-T/4t)\sqrt{t\ln \left[\frac{\lambda T(1+dT/4t)}{(4\pi
t)^{d/2}}\right]}
\end{eqnarray}
and in accordance with Eqs. (50), (51) for the island collapse
point $r_{f}(t_{c})=0$ we find
\begin{eqnarray}
t_{c}=\frac{(\lambda
T)^{2/d}[1+2\pi/(\lambda^{2}T^{2-d})^{1/d}+\cdots]}{4\pi}.
\end{eqnarray}
Neglecting in Eq. (54) the terms ${\cal O}(T/t)$ we come exactly
to the long-time front trajectory asymptotics of the initially
uniform island [Eq.(13)]
$$
r_{f}=\sqrt{2dt\ln(t_{c}/t)},
$$
where now
$$
t_{c}=(N_{T})^{2/d}/4\pi,
$$
whence it follows that in the island maximal expansion point
$$
t_{c}/t_{M}=e,
$$
\begin{eqnarray}
r_{f}^{M}=\sqrt{2dt_{M}}=(N_{T})^{1/d}\sqrt{d/2\pi e}.
\end{eqnarray}
Thus, we conclude that as well as in the case of the initially
uniform island, i.e. regardless of the initial $A$-particles
distribution, the long-time front trajectory of the long-living
island regardless of the system dimension is described by the
universal law (16)
$$
\zeta_{f}= r_{f}/r_{f}^{M}=\sqrt{e\tau|\ln\tau|}, \quad
\tau=t/t_{c}.
$$
Satisfying within the island the requirement $r^{2}T/t^{2}\leq
r_{f}^{2}T/t^{2}\ll 1$ we find the key condition for front
trajectory universalization
\begin{eqnarray}
\tau \gg (T/t_{c}){\rm max}(1,|\ln\tau|).
\end{eqnarray}
As an illustration, Figs. 4a and 4b demonstrate collapse of 1D and
3D islands front trajectories to the universal trajectory (16)
with the growing parameters $\lambda ^{2}T$ and $\lambda ^{2}/T$,
respectively.

\begin{figure}
\includegraphics[width=1\columnwidth]{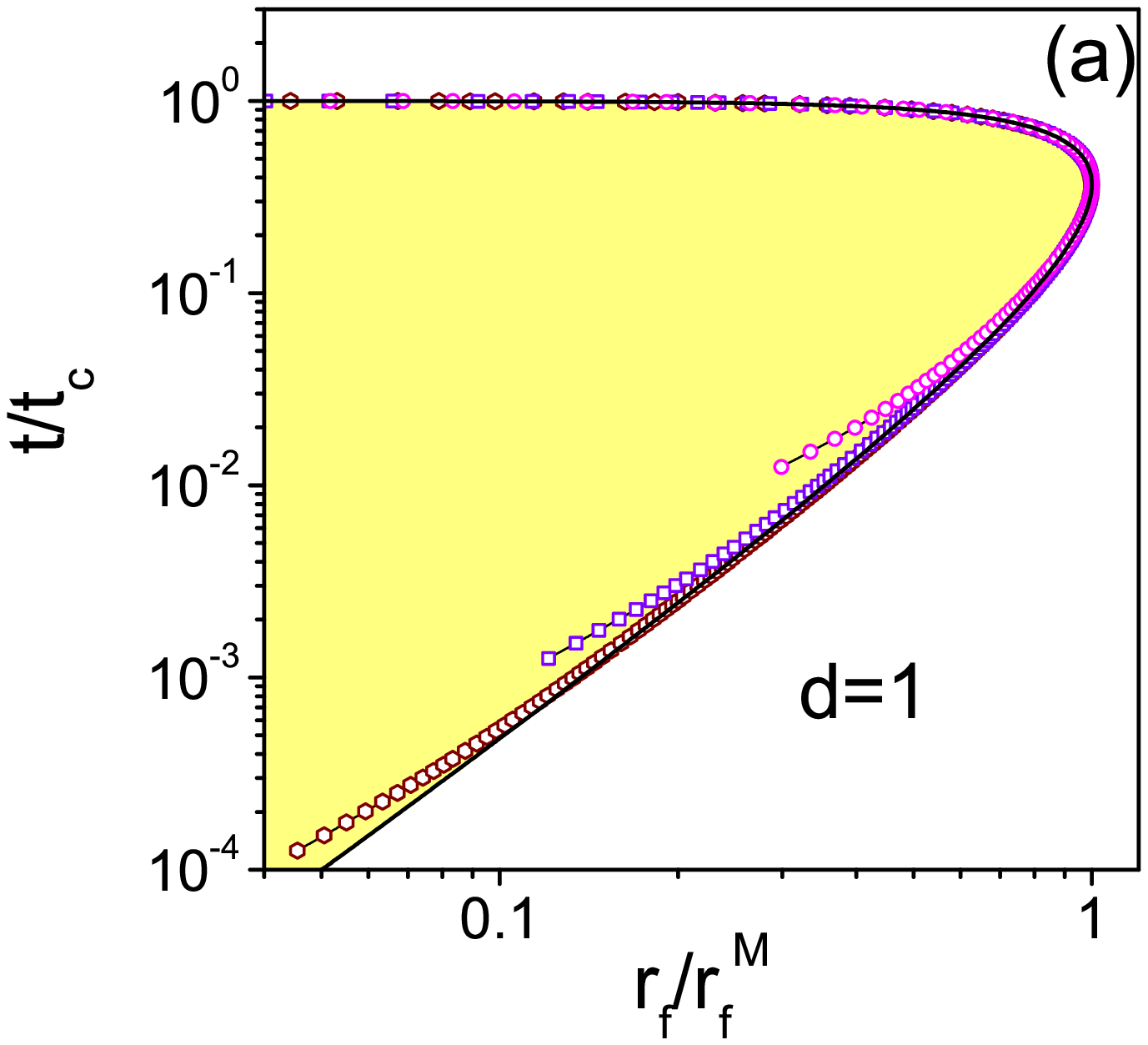}
\label{fig 4a}
\end{figure}
\begin{figure}
\includegraphics[width=1\columnwidth]{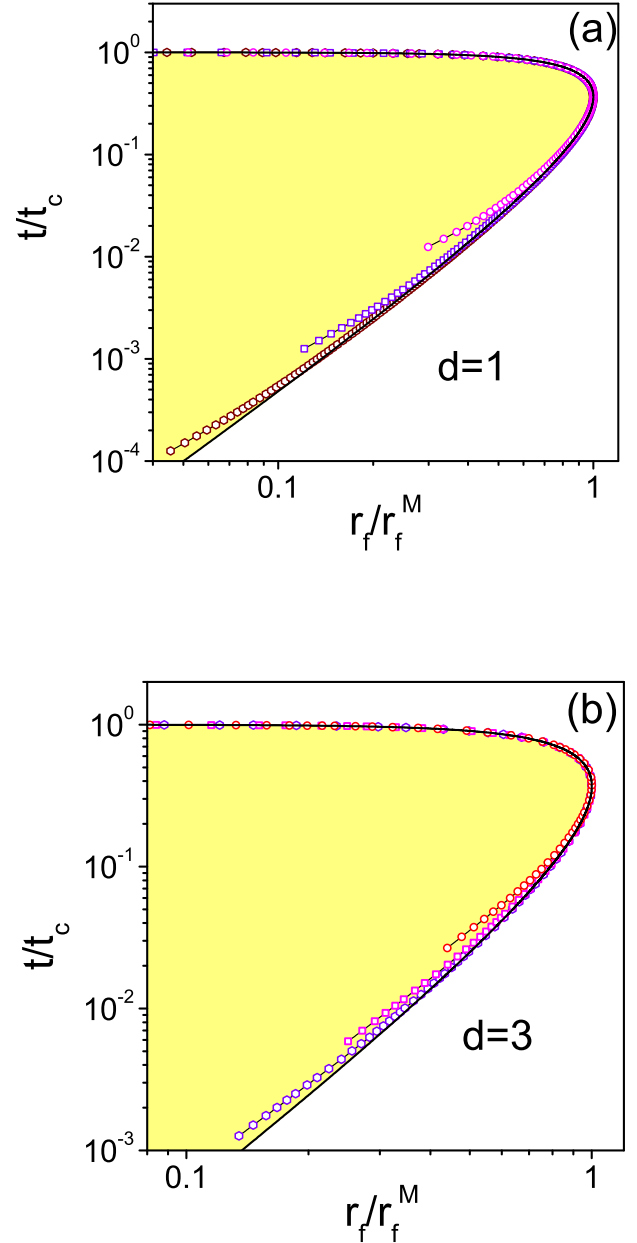}
\caption {(Color online) Collapse of the shown in figures 2 and 3
front trajectories after the source switching-off to the universal
trajectory $\zeta_{f}(\tau)$ (16)(thick line) in the scaling
coordinates $r_{f}/r_{f}^{M}$ vs $t/t_{c}$ with growing parameters
$\lambda^{2}T$ and $\lambda^{2}/T$, respectively: (a) $d=1$,
$\lambda^{2}T=10^{3}$(circles), $10^{4}$(squares) and $10^{5}$
(hexagons); (b) $d=3$, $\lambda^{2}/T=10^{8}$(circles),
$10^{10}$(squares) and $10^{12}$ (hexagons).}
\label{fig 4}
\end{figure}

Assuming that the condition (57) is fulfilled and neglecting in
Eq. (53) the terms $(d-r^{2}/2t)T/4t+\cdots $, as well as in the
case of the initially uniform island we come to the independent of
the starting distribution self-similar evolution of the long-time
particles distribution in the island $a(\zeta ,\tau )$[Eq. 17]
and, as a consequence, we obtain the scaling laws of particles
death in the form
\begin{eqnarray}
N/N_{T}={\cal G}_{d}({\tau}),
\end{eqnarray}
where scaling functions ${\cal G}_{d}({\tau })$ are determined by
the expression (18). In the limit of large $t_{c}/T\rightarrow
\infty $ hence it follows that in the front turning point $\tau
_{M}$ {\it regardless of the number of the injected particles}
$N_{T}=\lambda T$ the particles distribution in the island takes
the form
$$
a(\zeta, \tau_{M})=e^{d(1-\zeta^{2})/2}-1,
$$
and the fraction of the particles survived at the island expansion
stage $N_{M}/N_{T}$ is determined by the expression (20). From Eq.
(58) it also follows that at the final island collapse stage the
fraction of the surviving particles $N({\cal T})/N_{T}$ decays
according to the power law (21). Completing the long-living island
evolution analysis it should be emphasized that according to Eqs.
(16), (17) and (57) as well as in the case of the initially
uniform island we conclude that regardless of the system dimension
the reduced boundary current behavior ${\cal J}(\tau )=J/J_{M}$ is
described by the universal law (22) that predetermines
universality of the mean-field front relative width evolution.

\section{Evolution of the reaction front}

So far we have assumed formally that the reaction front is sharp
enough so that the front relative width $\eta =w/r_{f}$ remains
negligibly small up to a narrow vicinity of the island collapse
point. In this section we shall reveal the conditions for this
assumption realization in the long-living island regime. Supposing
that $|\zeta -\zeta _{f}|/\zeta _{f}\ll 1$ we find from Eqs. (16)
and (17)
$$
s+1=e^{-d|\ln \tau|(\zeta-\zeta_{f})/\zeta_{f}+\cdots},
$$
whence it follows that in the front center vicinity $|\zeta -\zeta
_{f}|/\zeta _{f}\ll {\rm min}(1,1/d|\ln \tau |)$ the quantity $s$
becomes a linear function of $\zeta $
$$
s = -d|\ln \tau|(\zeta-\zeta_{f})/\zeta_{f}+ \cdots.
$$
It means that the boundary current density (22) determines the
evolution of the quasistatic reaction front width $w(J)$ in
fulfilling the requirement
\begin{eqnarray}
\eta = w/r_{f}\ll {\rm min}(1,1/d|\ln\tau|).
\end{eqnarray}
It is not difficult to show that in accordance with Eq. (22) the
front quasistaticity condition $t_{w}/t_{J}\ll 1$ (where the value
$t_{J}=-(d\ln J/dt)^{-1}$ describes the rate of the boundary
current change and $t_{w}\sim w^{2}$ is the equilibration time of
the reaction front) takes the form
$$
t_{w}/t_{J}\sim d\eta^{2}(1+|\ln\tau|)\ll 1,
$$
whence it follows self-consistently that in fulfilling the
requirement (59) the reaction front becomes quasistatic
automatically.

In the works \cite{cor}- \cite{koza} it is established that at
$d>d_{c}=2$ in the dimensional variables the dependence of the
quasistatic front width on the boundary current density is
described by the mean-field law
\begin{eqnarray}
w_{\rm MF}\sim (D^{2}/kJ)^{1/3},
\end{eqnarray}
whereas in the 1D case in the diffusion-controlled limit the
quasistatic front width becomes $k$-independent and it is
determined by the fluctuation law
\begin{eqnarray}
w_{\rm F}\sim \sqrt{D/J}.
\end{eqnarray}
At upper critical dimension $d=d_{c}=2$ in the
diffusion-controlled limit in the mean-field law (60) a
logarithmic correction appears (logarithmically modified front)
$w_{{\rm L}}\propto (|\ln J|/J)^{1/3}$ \cite{lee}, \cite{hov},
\cite{kra}; its full form will be presented below. At first we
shall consider the evolution of the 1D fluctuation front width,
then we shall analyze the behavior of the modified two-dimensional
front width and finally we shall reveal the regularities of the
mean-field front width evolution for quasi-one-dimensional,
quasi-two-dimensional and three-dimensional geometry.

\subsection{Fluctuation front}

According to Eq. (61) in the units that we have accepted the
fluctuation front width reads
\begin{eqnarray}
\nonumber
w_{\rm F}\sim 1/\sqrt{n_{0}J},
\end{eqnarray}
where $n_{0}=b_{0}L$ for the initially uniform island and
$n_{0}=\rho \ell =1$ for the island formed by the localized
source. Substituting here Eq. (22) we find
\begin{eqnarray}
w_{\rm F}= w_{\rm F}^{M}(e\tau/|\ln\tau|)^{1/4},
\end{eqnarray}
where $w_{\rm F}^{M}\sim \sqrt{r_{f}^{M}/n_{0}}$. From Eqs. (62)
and (16) it follows that the fluctuation front relative width
$\eta _{{\rm F}}=w_{{\rm F}}/r_{f}$ changes by the law
\begin{eqnarray}
\eta_{\rm F}= \eta_{\rm F}^{M}/(e\tau|\ln^{3}\tau|)^{1/4},
\end{eqnarray}
where in accordance with Eqs. (15) and (56) the relative width
amplitude in the front turning point is
\begin{eqnarray}
\eta_{\rm F}^{M}\sim 1/\sqrt{n_{0}r_{f}^{M}}\sim 1/\sqrt{\cal N}
\end{eqnarray}
and ${\cal N}$ is the initial number of particles in the
originally uniform island ${\cal N}=n_{0}N_{0}$ or the number of
injected particles ${\cal N}=N_{T}$. According to Eq. (63) at the
island expansion stage the value $\eta _{{\rm F}}$ is decreasing
relatively slowly reaching the minimum ${\rm min}(\eta _{{\rm
F}})\approx 0.72\eta _{{\rm F}}^{M}$ at $\tau _{m}=1/e^{3}$ and
then, at the island contraction stage at ${\cal T}=1-\tau
=(t_{c}-t)/t_{c}\ll 1$ it begins to increase fast by the law
$$
\eta_{\rm F}\sim ({\cal T}_{Q}/{\cal T})^{3/4},
$$
where ${\cal T}_{Q}\sim 1/{\cal N}^{2/3}\to 0$ as ${\cal
N}\to\infty$. Thus, we conclude that at sufficiently large initial
number of particles ${\cal N}$ the fluctuation front remains sharp
enough up to a narrow vicinity of the island collapse point.
According to Eqs. (59),(63) and (64) at the island expansion stage
far from the collapse point ($\tau \ll 1$) the front becomes sharp
$\eta _{{\rm F}}|\ln \tau |<\epsilon \ll 1$ under the condition
$$
\tau/|\ln\tau|> 1/(\epsilon^{2}{\cal N})^{2}.
$$
It should be emphasized that in contrast to the front width the
amplitude $\eta _{{\rm F}}^{M}$ determining characteristic scale
of the fluctuation front relative width does not depend on the sea
density.

\subsection{Modified front at $d=d_{c}=2$}

Following Krapivsky's approach \cite{kra} along with the arguments
of Ref. \cite{ben}, for the modified two-dimensional front width
in the diffusion-controlled limit we find (in dimensional
variables)
\begin{eqnarray}
\nonumber w_{\rm L}\sim
\left[\left(\frac{D}{J}\right)\ln\left(\frac{D}{Jr_{a}^{3}}\right)\right]^{1/3},
\end{eqnarray}
where $r_{a}$ is the reaction radius and $D/Jr_{a}^{3}\gg 1$
according to the requirement $w_{{\rm L}}\gg r_{a}$. Substituting
here Eq. (22) we obtain
\begin{eqnarray}
w_{\rm L}=w_{\rm
L}^{M}\frac{[1+\ln(\sqrt{e\tau/|\ln\tau|})/\ln\phi]^{1/3}}{(|\ln\tau|/e\tau)^{1/6}},
\end{eqnarray}
whence in accordance with Eqs. (15), (16) and (56) it follows that
the front relative width $\eta _{{\rm L}}=w_{{\rm L}}/r_{f}$
changes by the law
\begin{eqnarray}
\eta_{\rm L}=\eta_{\rm
L}^{M}\left[\frac{1+\ln(\sqrt{e\tau/|\ln\tau|})/\ln
\phi}{e\tau\ln^{2}\tau}\right]^{1/3},
\end{eqnarray}
where the relative width amplitude in the front turning point is
\begin{eqnarray}
\eta_{\rm L}^{M}\sim \left[\frac{\ln\phi({\cal N})}{{\cal
N}}\right]^{1/3},
\end{eqnarray}
$$
\phi({\cal N})=(\ell/r_{a})^{3}\sqrt{\cal N},
$$
and ${\cal N}$ is the initial number of particles in the
originally uniform island ${\cal N}=b_{0}L^{2}N_{0}$ or the number
of injected particles ${\cal N}=N_{T} $. According to Eq. (66) at
the island expansion stage the value $\eta _{{\rm L}}$ is
decreasing relatively slowly reaching at large enough $\phi $ the
minimum ${\rm min}(\eta _{{\rm L}})\approx 0.88(1-0.28/\ln \phi
+\cdots )\eta _{{\rm L}}^{M}$ at $\tau _{m}=e^{-2}(1-3/2\ln \phi
+\cdots )$ and then, at the island contraction stage at ${\cal
T}\ll 1$ it begins to increase fast by the law
$$
\eta_{\rm L}\sim ({\cal T}_{Q}/{\cal T})^{2/3},
$$
where ${\cal T}_{Q}\sim [\ln(\phi/\sqrt{\cal T})/{\cal
N}]^{1/2}\to 0$ as ${\cal N}\to\infty$ at fixed ${\cal T}$. From
Eqs. (59) and (66) it follows that at the island expansion stage
far from the collapse point ($\tau \ll 1$) the front becomes sharp
$\eta _{{\rm L}}|\ln \tau |<\epsilon \ll 1$ under the condition
$$
\tau/|\ln\tau|>
\frac{\ln(\phi\sqrt{e\tau/|\ln\tau|})}{\epsilon^{3}\cal N}
$$
with the additional requirement $\tau/|\ln\tau|\gg \phi^{-2}$.
Thus, we conclude that at sufficiently large "starting" number of
particles ${\cal N}$ the modified two-dimensional front becomes
sharp at early stages of the island expansion and remains sharp up
to a narrow vicinity of the collapse point. Let us note that in
contrast to the fluctuation front the relative width amplitude
$\eta _{{\rm L}}^{M}$ increases logarithmically slowly at the
decrease of the sea density. As an illustration, assuming that
$r_{a}\sim 10^{-8}cm,b_{0}=10^{12}cm^{-2},L=0.1cm,$ and $c=10^{3}$
we find ${\cal N}\approx 3\times 10^{13}$, $\ln \phi \approx 30$
and $\eta _{{\rm L}}^{M}\approx 10^{-4}$.

\subsection{Mean-field front}

According to Eq. (60) in the units that we have accepted the
mean-field front width reads
\begin{eqnarray}
\nonumber
w_{\rm MF}\sim 1/(\kappa  J)^{1/3},
\end{eqnarray}
where $\kappa =kb_{0}L^{2}/D$ for the originally uniform island
and $\kappa =k\rho \ell ^{2}/D$ for the island formed by the
localized source. Substituting here Eq. (22) we find
\begin{eqnarray}
w_{\rm MF}= w_{\rm MF}^{M}(e\tau/|\ln\tau|)^{1/6},
\end{eqnarray}
where $w_{\rm MF}^{M}\sim({r_{f}^{M}/d\kappa})^{1/3}$. From Eqs.
(68) and (16) it follows that the relative mean-field front width
$\eta _{{\rm MF}}=w_{{\rm MF}}/r_{f}$ changes by the law
\begin{eqnarray}
\eta_{\rm MF}= \eta_{\rm MF}^{M}/(e\tau \ln^{2}\tau)^{1/3},
\end{eqnarray}
where in accordance with Eqs. (15) and (56) the relative width
amplitude in the front turning point is
\begin{eqnarray}
\eta_{\rm MF}^{M}\sim (\sqrt{d\kappa}r_{f}^{M})^{-2/3}=
m\left(\frac{D\ell^{d-2}}{k}\right)^{1/3}{\cal N}^{-2/3d},
\end{eqnarray}
where $m=(2\pi e/d^{2})^{1/3}$ and ${\cal N}$ is the initial
number of particles in the originally uniform island ${\cal
N}=b_{0}L^{d}N_{0}$ or the number of the injected particles ${\cal
N}=N_{T}$. According to Eq. (69) at the island expansion stage the
value $\eta _{{\rm MF}}$ is decreasing relatively slowly reaching
the minimum ${\rm min}(\eta _{{\rm MF}})\approx 0.88\eta _{{\rm
MF}}^{M}$ at $\tau _{m}=1/e^{2}$ and then, at the island
contraction stage at ${\cal T}\ll 1$ it begins to increase fast by
the law
$$
\eta_{\rm MF}\sim ({\cal T}_{Q}/{\cal T})^{2/3},
$$
where ${\cal T}_{Q}\sim (\eta_{\rm MF}^{M})^{3/2}\propto {\cal
N}^{-1/d}\to 0 $ as ${\cal N}\to\infty$. From Eqs. (59) and (69)
it follows that at the island expansion stage far from the
collapse point ($\tau \ll 1$) the front becomes sharp $\eta _{{\rm
MF}}|\ln \tau |<\epsilon \ll 1$ under the condition
$$
\tau/|\ln\tau|>(\eta_{\rm MF}^{M}/\epsilon)^{3}\propto
\epsilon^{-3}{\cal N}^{-2/d}.
$$
Barkema, Howard and Cardy have shown analytically and numerically
\cite {bar} that in the 1D case the fluctuation front is formed
under the condition $w_{{\rm F}}/w_{{\rm MF}}\gg 1$
$(k/\sqrt{JD}\gg 1)$ while in the opposite limit $w_{{\rm
F}}/w_{{\rm MF}}\ll 1$ $(k/\sqrt{JD}\ll 1)$ the front width is
determined by the mean-field law (60). Comparing Eqs. (63) and
(69) we find $\eta _{{\rm F}}/\eta _{{\rm MF}}=(\eta _{{\rm
F}}^{M}/\eta _{{\rm MF}}^{M})(e\tau /\ln \tau )^{1/12}$ whence by
virtue of weak dependence on $\tau $ it follows that as a
characteristic crossover point from the mean-field to the
fluctuation regime $({\rm MF}\rightarrow {\rm F}) $ it is
reasonable to accept the ratio $\eta _{{\rm F}}^{M}/\eta _{{\rm
MF}}^{M}\sim 1$. Substituting here Eqs. (64) and (70 ) for the
fluctuation regime area we find
$$
k\gg k_{\rm F}\sim D/\ell\sqrt{\cal N},
$$
whence it follows that with the increase in the initial number of
particles in the island and the decrease of the sea density the
fluctuation regime domain expands indefinitely ($k_{{\rm
F}}\rightarrow 0$ as $\ell \sqrt{{\cal N}}\rightarrow \infty $).
Determining further the lower bound of the sharp mean-field front
regime by the condition $\eta _{{\rm MF}}^{M}<\varepsilon \ll 1$
we find from Eq. (70)
$$
k> k_{\rm MF}^{\varepsilon}\sim (D/\ell)/\varepsilon^{3}{\cal
N}^{2},
$$
whence it follows
$$
 k_{\rm F}/k_{\rm MF}^{\varepsilon}\sim (\varepsilon^{2}{\cal
 N})^{3/2}
$$
and we conclude that the area of the island death in the sharp
mean-field front regime $k_{{\rm MF}}^{\varepsilon }< k\ll k_{{\rm
F}}$ appears under the condition ${\cal N}>\varepsilon ^{-2}(\eta
_{{\rm F}}^{M}<\varepsilon )$ and expands fast with the increase
of ${\cal N}$.

Repeating the presented above argumentation, for the crossover
from the mean-field to the logarithmically modified front (${\rm
MF}\rightarrow {\rm L})$ at $d=d_{c}=2$ we find from Eqs.(67) and
(70)
$$
k\gg k_{\rm L}\sim D/\ln\phi({\cal N}),
$$
whence it follows that with the increase of the starting number of
the particles in the island and the decrease of the sea density
the LM regime area expands logarithmically slowly. Determining the
lower bound of the sharp mean-field front regime by the condition
$\eta _{{\rm MF}}^{M}<\varepsilon \ll 1$ we find from Eq. (70)
$$
k >k_{\rm MF}^{\varepsilon}\sim D/{\cal N}\varepsilon^{3},
$$
whence it follows
$$
k_{\rm L}/k_{\rm MF}^{\varepsilon}\sim \varepsilon^{3} {\cal
 N}/\ln\phi({\cal N})
$$
and we conclude that the area of the island death in the sharp
mean-field front regime $k_{{\rm MF}}^{\varepsilon }<k\ll k_{{\rm
L}}$ appears under the condition ${\cal N}/\ln \phi
>\varepsilon ^{-3}(\eta _{{\rm L}}^{M}<\varepsilon )$ and expands
fast with the increase of ${\cal N}$.

According to Eq. (70) in the 3D case with the growth of the
reaction constant the front relative width amplitude decreases
$\propto k^{-1/3}$ reaching the minimal value in the
diffusion-controlled limit of the perfect reaction $k=k_{p}=8\pi
Dr_{a}$ where $r_{a}$ is the reaction radius. Substituting $k_{p}$
into Eq. (70) we find
\begin{eqnarray}
\nonumber
\eta_{\rm MF}^{M}\sim
m_{p}\left(\frac{\ell}{r_{a}}\right)^{1/3}{\cal N}^{-2/9}
\end{eqnarray}
with $m_{p}\approx 0.4$ whence taking for illustration $r_{a}\sim
10^{-8}cm,b_{0}=10^{20}cm^{-3},L=0.1cm,$ and $c=10^{3}$ we obtain
$\eta _{{\rm MF}}^{M}\approx 3\times 10^{-5}$. Thus, we conclude
that the three-dimensional spherical island dies in the sharp
front regime in a wide range of parameters (for instance at the
decrease of the reaction constant by 9 orders of magnitude the
front keeps sharp enough). It should be emphasized that according
to Eq. (70) in the region of the sharp mean-field front existence
at the decrease of the sea density the front relative width
amplitude decreases in the 1D case, does not change in the 2D case
and increases in the 3D case.

\subsection{Quasi-$d$-dimensional systems}

So far we have analyzed the evolution of the $d$-dimensional
spherical island formed by either the point source or the
initially uniform spherically symmetric particles distribution.
Completing this section we shall consider for completeness the
island evolution in the three-dimensional medium for the
quasi-one-dimensional (planar front) and the quasi-two-dimensional
(cylindrical front) geometry. Let in the uniform three-dimensional
$B$-particle sea acts a) a planar two-dimensional source of $A$
particles with the injection rate $\Lambda _{+}$ particles in a
time unit per a source unit area or b) a linear one-dimensional
source with the injection rate $\Lambda _{+}$ particles in a time
unit per a source unit length. Then, by virtue of symmetry, a
"planar" island with a width $2x_{f}(t)=2r_{f}(t)$ (wherein the
concentration changes only along the normal to the source plane)
will be formed around the planar source, and a cylindrical island
with a radius $r_{f}(t)$ (wherein the concentration changes only
along the normal to the source axis) will be formed around the
linear source. It is not difficult to show that in the units that
we have accepted all the results obtained in Sect. III remain
valid for the effective dimension $d_{+}$ with the only difference
that now the reduced source strength takes the form
$$
\lambda = \Lambda_{+}\ell^{2+\delta}/D,
$$
where $\delta =3-d_{+}$ and the effective dimension of the system
is $d_{+}=1(\delta =2)$ for the planar source and $d_{+}=2(\delta
=1)$ for the linear source (nevertheless as before $\ell =\rho
^{-1/d}=\rho ^{-1/3})$. Besides it is clear that instead of the
number of particles in the island $N$ in the quasi-one-dimensional
and the quasi-two-dimensional geometry the reduced number of
particles appears
$$
N_{+}\ell^{\delta}= g_{d_{+}}\int_{0}^{r_{f}}s(r,t)r^{d_{+}-1}dr,
$$
where $N_{+}$ is the number of particles in the island per unit
area $(d_{+}=1)$ or per unit length $(d_{+}=2)$ of the source, and
the reduced number of injected particles is
$$
N_{T}=\lambda T = {\cal N}_{+}\ell^{\delta},
$$
where ${\cal N}_{+}$ is the number of injected particles per unit
area $(d_{+}=1)$ or per unit length $(d_{+}=2)$ of the source. In
the long-living island regime for the island radius amplitude in
the front turning point we find from Eq. (56)
\begin{eqnarray}
\nonumber
r_{f}^{M}=\sqrt{\frac{d_{+}}{2\pi e}}({\cal
N}_{+}\ell^{\delta})^{1/d_{+}},
\end{eqnarray}
whence after substitution to Eq. (70) for the front relative width
amplitude we obtain
\begin{eqnarray}
\eta_{\rm MF}^{M}\sim
m_{+}\left(\frac{D\ell^{\sigma}}{k}\right)^{1/3}{\cal
N_{+}}^{-2/3d_{+}},
\end{eqnarray}
where $\sigma=3(d_{+}-2)/d_{+}$ and $m_{+}=(2\pi
e/d_{+}^{2})^{1/3}$.

Let now the uniform "planar" $A$-particle island with a width $2L$
or the uniform cylindrical $A$-particle island with a radius $L$
and the initial concentration $a_{0}=cb_{0}$ ($c\gg 1)$ be
surrounded by the uniform three-dimensional $B$-particle sea with
the concentration $b_{0}$. Then, by virtue of symmetry, as well as
in the case with the planar and the linear sources, in the course
of the following evolution a "planar" island should remain the
"planar" one wherein the concentration changes only along the
normal to the front plane $(d_{+}=1)$, and a cylindrical island
should remain the cylindrical one wherein the concentration
changes only along the normal to the cylinder axis $(d_{+}=2)$. It
is not difficult to show that in the units that we have accepted
all the results obtained in Sect. II remain valid for the
effective dimension $d_{+}$ with the only difference that instead
of the reduced number of particles in the island $N$ evaluated in
the units $b_{0}L^{d}$, in the quasi-one-dimensional and the
quasi-two-dimensional systems there appears the reduced number of
particles in the island $N_{+}$ per a front unit area $(d_{+}=1)$
or per a cylinder axis unit length $(d_{+}=2)$ evaluated in the
units $b_{0}L^{d_{+}}$ with the initial number of particles in the
island ${\cal N}_{+}=N_{0}b_{0}L^{d_{+}}=\mu
_{d_{+}}cb_{0}L^{d_{+}}$ per a unit of area and length,
respectively. In the long-living island regime for the island
radius amplitude in the front turning point we find from Eq. (15)
\begin{eqnarray}
\nonumber
r_{f}^{M}=\sqrt{\frac{d_{+}}{2\pi e}}(N_{0})^{1/d_{+}},
\end{eqnarray}
whence after substitution to Eq. (70) taking into account the
relation $\ell =b_{0}^{-1/3}$ we come to Eq. (71) again for the
front relative width amplitude. Substituting further to Eq. (71)
the constant of the diffusion-controlled perfect reaction
$k=k_{p}=8\pi Dr_{a}$ we find
$$
\eta_{\rm MF}^{M}\sim
m_{p+}\left(\frac{\ell^{\sigma}}{r_{a}}\right)^{1/3}{\cal
N_{+}}^{-2/3d_{+}}
$$
with $m_{p+}=(e/4d_{+}^{2})^{1/3}$ whence taking for illustration
the same parameters as for the spherical island $r_{a}\sim
10^{-8}cm,b_{0}=10^{20}cm^{-3},L=0.1cm,$ and $c=10^{3}$ we obtain
${\cal N}_{+}\approx 2\times 10^{22}cm^{-2}$, $\eta _{{\rm
MF}}^{M}\approx 3\times 10^{-6}$ for the quasi-one-dimensional
island $(d_{+}=1)$ and ${\cal N}_{+}\approx 3\times
10^{21}cm^{-1}$, $\eta _{{\rm MF}}^{M}\approx 2\times 10^{-5}$ for
the quasi-two-dimensional island $(d_{+}=2)$. Thus, we conclude
that at enough large ${\cal N}_{+}$ the quasi-one-dimensional and
the quasi-two-dimensional islands die in the three-dimensional sea
in the sharp front regime in a wide interval of parameters. One
can easily be convinced that this conclusion remains valid for the
quasi-one-dimensional geometry in the two-dimensional medium
(linear one-dimensional source in the two-dimensional sea with the
logarithmically modified front).

\section{Conclusion}

In this paper, we have presented a systematic analytical study of
diffusion-controlled formation and collapse of a $d$-dimensional
$A$-particle island in the $B$-particle sea at propagation of the
sharp reaction front $A+B\to 0$. Our main purpose was to describe
the formation regularities for the $d$-dimensional spherical
island at $A$ particles injection by a point source acting for
some finite time $T$ and the following island evolution and
collapse after source switching-off. We have focused mainly on the
most interesting case when the island collapse time $t_{c}$
becomes much longer than the injection period $T$ (long-living
island) and have revealed the complete picture of the evolution of
front trajectory and particle distribution in the island depending
on the intensity and duration of source action. Generalizing the
results obtained earlier for the quasi-one-dimensional geometry we
have also revealed the long-time evolution regularities for the
initially uniform $d$-dimensional spherical $A$-particle island.
The main results can be formulated as follows:

(1) The conditions of the long-living island formation have been
found and it was shown that in the long-living island regime the
ratio $t_{c}/T$ changes with the increase of the injection period
$T$ and the reduced source strength $\lambda $ by the law $\propto
(\lambda ^{2}T^{2-d})^{1/d}$, i.e. it increases with the increase
of $T$ in the 1D case, does not change with the increase of $T$ in
the 2D case and decreases with the increase of $T$ in the 3D case.
It has been established that {\it regardless of the medium
dimension and the injection duration} the ratio of the maximal
$d$-dimensional island volume in the front turning point $\Omega
_{M}$ to the initial island volume at the moment of source
switching-off $\Omega _{T}$ is proportional to the mean
concentration of the island particles at the moment of source
switching-off and is inversely proportional to the fraction of the
particles died by this moment.

(2) It has been established that {\it regardless of the number of
the injected particles and the system dimension} the long-time
front trajectory of the long-living island is described by the
universal law
$$
\zeta_{f}= r_{f}/r_{f}^{M}=\sqrt{e\tau|\ln\tau|},
$$
where $\tau=t/t_{c}$ and $r_{f}^{M}=(N_{T})^{1/d}\sqrt{d/2\pi e}$
is the radius of the island maximal expansion at the front turning
point $t_{M}=t_{c}/e$. The scaling laws of evolution of the
distribution and the number of particles in the $d$-dimensional
long-living island have been derived and it has been shown that
{\it regardless of the system dimension} the evolution of the
boundary current density $J$ that determines the quasistatic front
width is described by the universal law
$$
{\cal J}(\tau)=J/J_{M}=\sqrt{\frac{|\ln\tau|}{e\tau}}.
$$

(3) It has been shown that regardless of the initial particle
distribution the long-time evolution of the initially uniform
concentrated island as well as the long-time evolution of the
long-living island converge to the unified universal island death
asymptotics in the instantaneous source regime
$$
a(\zeta, \tau)=(e^{-\zeta^{2}/e\tau}/\tau)^{d/2} - 1.
$$

(4) The systematic analysis of the reaction front relative width
evolution for the fluctuation, the logarithmically modified and
the mean-field regimes was presented and it was demonstrated that
in a wide range of parameters at a large enough number of injected
or initially uniformly distributed particles the front remains
sharp up to a narrow vicinity of the island collapse point.

In conclusion it should be emphasized that as well as in the paper
\cite{self1}, here the evolution of the island has been considered
at equal species diffusivities. Although we believe that the
regularities discovered reflect the key features of the island
evolution the study of the much more complicated problem for
unequal species diffusivities remains a challenging problem for
the future.

\begin{acknowledgments}
This research was financially supported by the RFBR through Grant
No 08-03-00054. I am grateful to Dr. V.G. Rostiashvili for
valuable comments.
\end{acknowledgments}

%\caption {(Color online) Filled circles - trajectory of the front
%radius for the continuous in time point source in the
%one-dimensional medium calculated according to Eq. (30) in the
%scaling coordinates $\lambda r_{f}$ vs. $\lambda^{2}t$. Dashed
%line shows long-time asymptotics Eq.(37). Open squares -
%trajectories of the front radius after the source switching-off
%calculated from Eq. (44) for the parameter values $\lambda^{2}T =
%10^{2}, 10^{3}, 10^{4}$ and $10^{5}$(from bottom to top).}


\begin{references}
\bibitem{kbr} P.L. Krapivsky, E. Ben-Naim and S. Redner, {\it A
Kinetic View of Statistical Physics} (Cambridge University Press,
Cambridge, 2010)
\bibitem{rev} D. ben Avraham and S. Havlin, {\it Diffusion and Reactions in
Fractals and Disodered Systems} (Cambridge University Press,
Cambridge, 2000)
\bibitem {cho} B. Chopard and M. Droz, {\it Cellular
automata modelling of physical systems} (Cambridge University
Press, Cambridge, 1998)
\bibitem{cot} E. Kotomin and V. Kuzovkov, {\it
Modern Aspects of Diffusion Controlled Reactions: Cooperative
Phenomena in Bimolecular Processes} (Elsevier, Amsterdam, 1996).
\bibitem{ant} T. Antal, M. Droz, J. Magnin, and Z. Racz, Phys.
Rev. Lett., {\bf 83}, 2880 (1999).
\bibitem{rz} Z. Racz, Physica A, {\bf 274}, 50, (1999)
\bibitem{tho} S. Thomas, I. Lagzi, F. Molnar, Jr., and Z. Racz, Phys.
Rev. Lett., {\bf 110}, 078303 (2013).
\bibitem{but1} L.V. Butov, A.C. Gossard and D.S. Chemla, Nature, {\bf 418 }, 751 (2002)
\bibitem{snoke} D. Snoke, S. Denev, Y. Lin, L. Pfeiffer and K.
West, Nature, {\bf 418 }, 754 (2002)
\bibitem{but2} Sen Yang, L.V. Butov, L.S. Levitov, B.D. Simons and
A.C. Gossard, Phys. Rev. B {\bf 80}, 155331 (2009)
\bibitem{gal} L. Galfi and Z. Racz, Phys. Rev. A {\bf 38}, 3151 (1988).
\bibitem{ben} E. Ben-Naim and S. Redner, J. Phys. A {\bf 28}, L575 (1992).
\bibitem{cor} S. Cornell and M. Droz, Phys. Rev. Lett. {\bf 70}, 3824 (1993).
\bibitem{lee} B.P. Lee and J. Cardy, Phys. Rev. E  {\bf 50}, R3287 (1994).
%\bibitem{cor2} S. Cornell, Phys. Rev. Lett. {\bf 75}, 2250 (1995).
%\bibitem{ben} E. Ben-Naim and S. Redner, J. Phys. A {\bf 28}, L575 (1992).
\bibitem{bar} G.T. Barkema, M.J. Howard and J.L. Cardy, Phys. Rev. E {\bf 53}, R2017 (1996).
\bibitem{koza} Z. Koza, J. Stat. Phys. {\bf 85}, 179 (1996).
\bibitem{self1} B.M. Shipilevsky, Phys. Rev. E {\bf 67}, 060101(R) (2003)
\bibitem{self2} B.M. Shipilevsky, Phys. Rev. E {\bf 70}, 032102 (2004)
\bibitem{self3} B.M. Shipilevsky, Phys. Rev. E {\bf 77}, 030101(R)(2008)
\bibitem{kis} S. Kisilevich, M. Sinder, J. Pelleg, and V.
Sokolovsky, Phys. Rev. E {\bf 77}, 046103 (2008)
\bibitem{self4} B.M. Shipilevsky, Phys. Rev. E {\bf 79}, 061114 (2009)
\bibitem{self5} B.M. Shipilevsky, Phys. Rev. E {\bf 82}, 011119 (2010)
\bibitem{hay} C.P. Haynes, R. Voituriez, and O. Benichou, J. Phys. A {\bf 45}, 415001 (2012)
\bibitem{self6} B.M. Shipilevsky, Phys. Rev. E {\bf 88 }, 012133 (2013)
\bibitem{fi} M. Fialkowski, A. Bitner, and B.A. Grzybowski, Phys.
Rev. Lett., {\bf 94}, 018303 (2005).
\bibitem{racz2} I. Lagzi, P. Papai and Z. Racz, Chem. Phys. Lett. {\bf 433 }, 286 (2007).
\bibitem{wi} N. Withers, Nature Chemistry, {\bf 2}, 160 (2010).
\bibitem{hov} M. Howard and J. Cardy, J. Phys. A {\bf 28}, 3599 (1995).
\bibitem{kra} P.L. Krapivsky, Phys. Rev. E {\bf 51}, 4774 (1995).
\end{references}
\end{document}